\documentstyle[12pt]{article}
\begin{document}

\immediate\write16{<<WARNING: LINEDRAW macros work with emTeX-dvivers
		    and other drivers supporting emTeX \special's
		    (dviscr, dvihplj, dvidot, dvips, dviwin, etc.) >>}

\newdimen\Lengthunit	   \Lengthunit	= 1.5cm
\newcount\Nhalfperiods	   \Nhalfperiods= 9
\newcount\magnitude	   \magnitude = 1000

\catcode`\*=11
\newdimen\L*   \newdimen\d*   \newdimen\d**
\newdimen\dm*  \newdimen\dd*  \newdimen\dt*
\newdimen\a*   \newdimen\b*   \newdimen\c*
\newdimen\a**  \newdimen\b**
\newdimen\xL*  \newdimen\yL*
\newdimen\rx*  \newdimen\ry*
\newdimen\tmp* \newdimen\linwid*

\newcount\k*   \newcount\l*   \newcount\m*
\newcount\k**  \newcount\l**  \newcount\m**
\newcount\n*   \newcount\dn*  \newcount\r*
\newcount\N*   \newcount\*one \newcount\*two  \*one=1 \*two=2
\newcount\*ths \*ths=1000
\newcount\angle*  \newcount\q*	\newcount\q**
\newcount\angle** \angle**=0
\newcount\sc*	  \sc*=0

\newtoks\cos*  \cos*={1}
\newtoks\sin*  \sin*={0}

\catcode`\[=13

\def\rotate(#1){\advance\angle**#1\angle*=\angle**
\q**=\angle*\ifnum\q**<0\q**=-\q**\fi
\ifnum\q**>360\q*=\angle*\divide\q*360\multiply\q*360\advance\angle*-\q*\fi
\ifnum\angle*<0\advance\angle*360\fi\q**=\angle*\divide\q**90\q**=\q**
\def\sgcos*{+}\def\sgsin*{+}\relax
\ifcase\q**\or
 \def\sgcos*{-}\def\sgsin*{+}\or
 \def\sgcos*{-}\def\sgsin*{-}\or
 \def\sgcos*{+}\def\sgsin*{-}\else\fi
\q*=\q**
\multiply\q*90\advance\angle*-\q*
\ifnum\angle*>45\sc*=1\angle*=-\angle*\advance\angle*90\else\sc*=0\fi
\def[##1,##2]{\ifnum\sc*=0\relax
\edef\cs*{\sgcos*.##1}\edef\sn*{\sgsin*.##2}\ifcase\q**\or
 \edef\cs*{\sgcos*.##2}\edef\sn*{\sgsin*.##1}\or
 \edef\cs*{\sgcos*.##1}\edef\sn*{\sgsin*.##2}\or
 \edef\cs*{\sgcos*.##2}\edef\sn*{\sgsin*.##1}\else\fi\else
\edef\cs*{\sgcos*.##2}\edef\sn*{\sgsin*.##1}\ifcase\q**\or
 \edef\cs*{\sgcos*.##1}\edef\sn*{\sgsin*.##2}\or
 \edef\cs*{\sgcos*.##2}\edef\sn*{\sgsin*.##1}\or
 \edef\cs*{\sgcos*.##1}\edef\sn*{\sgsin*.##2}\else\fi\fi
\cos*={\cs*}\sin*={\sn*}\global\edef\gcos*{\cs*}\global\edef\gsin*{\sn*}}\relax
\ifcase\angle*[9999,0]\or
[999,017]\or[999,034]\or[998,052]\or[997,069]\or[996,087]\or
[994,104]\or[992,121]\or[990,139]\or[987,156]\or[984,173]\or
[981,190]\or[978,207]\or[974,224]\or[970,241]\or[965,258]\or
[961,275]\or[956,292]\or[951,309]\or[945,325]\or[939,342]\or
[933,358]\or[927,374]\or[920,390]\or[913,406]\or[906,422]\or
[898,438]\or[891,453]\or[882,469]\or[874,484]\or[866,499]\or
[857,515]\or[848,529]\or[838,544]\or[829,559]\or[819,573]\or
[809,587]\or[798,601]\or[788,615]\or[777,629]\or[766,642]\or
[754,656]\or[743,669]\or[731,681]\or[719,694]\or[707,707]\or
\else[9999,0]\fi}

\catcode`\[=12

\def\GRAPH(hsize=#1)#2{\hbox to #1\Lengthunit{#2\hss}}

\def\Linewidth#1{\global\linwid*=#1\relax
\global\divide\linwid*10\global\multiply\linwid*\mag
\global\divide\linwid*100\special{em:linewidth \the\linwid*}}

\Linewidth{.4pt}
\def\sm*{\special{em:moveto}}
\def\sl*{\special{em:lineto}}
\let\moveto=\sm*
\let\lineto=\sl*
\newbox\spm*   \newbox\spl*
\setbox\spm*\hbox{\sm*}
\setbox\spl*\hbox{\sl*}

\def\mov#1(#2,#3)#4{\rlap{\L*=#1\Lengthunit
\xL*=#2\L* \yL*=#3\L*
\xL*=\xscale\xL* \yL*=\yscale\yL*
\rx* \the\cos*\xL* \tmp* \the\sin*\yL* \advance\rx*-\tmp*
\ry* \the\cos*\yL* \tmp* \the\sin*\xL* \advance\ry*\tmp*
\kern\rx*\raise\ry*\hbox{#4}}}

\def\rmov*(#1,#2)#3{\rlap{\xL*=#1\yL*=#2\relax
\rx* \the\cos*\xL* \tmp* \the\sin*\yL* \advance\rx*-\tmp*
\ry* \the\cos*\yL* \tmp* \the\sin*\xL* \advance\ry*\tmp*
\kern\rx*\raise\ry*\hbox{#3}}}

\def\lin#1(#2,#3){\rlap{\sm*\mov#1(#2,#3){\sl*}}}

\def\arr*(#1,#2,#3){\rmov*(#1\dd*,#1\dt*){\sm*
\rmov*(#2\dd*,#2\dt*){\rmov*(#3\dt*,-#3\dd*){\sl*}}\sm*
\rmov*(#2\dd*,#2\dt*){\rmov*(-#3\dt*,#3\dd*){\sl*}}}}

\def\arrow#1(#2,#3){\rlap{\lin#1(#2,#3)\mov#1(#2,#3){\relax
\d**=-.012\Lengthunit\dd*=#2\d**\dt*=#3\d**
\arr*(1,10,4)\arr*(3,8,4)\arr*(4.8,4.2,3)}}}

\def\arrlin#1(#2,#3){\rlap{\L*=#1\Lengthunit\L*=.5\L*
\lin#1(#2,#3)\rmov*(#2\L*,#3\L*){\arrow.1(#2,#3)}}}

\def\dasharrow#1(#2,#3){\rlap{{\Lengthunit=0.9\Lengthunit
\dashlin#1(#2,#3)\mov#1(#2,#3){\sm*}}\mov#1(#2,#3){\sl*
\d**=-.012\Lengthunit\dd*=#2\d**\dt*=#3\d**
\arr*(1,10,4)\arr*(3,8,4)\arr*(4.8,4.2,3)}}}

\def\clap#1{\hbox to 0pt{\hss #1\hss}}

\def\ind(#1,#2)#3{\rlap{\L*=.1\Lengthunit
\xL*=#1\L* \yL*=#2\L*
\rx* \the\cos*\xL* \tmp* \the\sin*\yL* \advance\rx*-\tmp*
\ry* \the\cos*\yL* \tmp* \the\sin*\xL* \advance\ry*\tmp*
\kern\rx*\raise\ry*\hbox{\lower2pt\clap{$#3$}}}}

\def\sh*(#1,#2)#3{\rlap{\dm*=\the\n*\d**
\xL*=\xscale\dm* \yL*=\yscale\dm* \xL*=#1\xL* \yL*=#2\yL*
\rx* \the\cos*\xL* \tmp* \the\sin*\yL* \advance\rx*-\tmp*
\ry* \the\cos*\yL* \tmp* \the\sin*\xL* \advance\ry*\tmp*
\kern\rx*\raise\ry*\hbox{#3}}}

\def\calcnum*#1(#2,#3){\a*=1000sp\b*=1000sp\a*=#2\a*\b*=#3\b*
\ifdim\a*<0pt\a*-\a*\fi\ifdim\b*<0pt\b*-\b*\fi
\ifdim\a*>\b*\c*=.96\a*\advance\c*.4\b*
\else\c*=.96\b*\advance\c*.4\a*\fi
\k*\a*\multiply\k*\k*\l*\b*\multiply\l*\l*
\m*\k*\advance\m*\l*\n*\c*\r*\n*\multiply\n*\n*
\dn*\m*\advance\dn*-\n*\divide\dn*2\divide\dn*\r*
\advance\r*\dn*
\c*=\the\Nhalfperiods5sp\c*=#1\c*\ifdim\c*<0pt\c*-\c*\fi
\multiply\c*\r*\N*\c*\divide\N*10000}

\def\dashlin#1(#2,#3){\rlap{\calcnum*#1(#2,#3)\relax
\d**=#1\Lengthunit\ifdim\d**<0pt\d**-\d**\fi
\divide\N*2\multiply\N*2\advance\N*\*one
\divide\d**\N*\sm*\n*\*one\sh*(#2,#3){\sl*}\loop
\advance\n*\*one\sh*(#2,#3){\sm*}\advance\n*\*one
\sh*(#2,#3){\sl*}\ifnum\n*<\N*\repeat}}

\def\dashdotlin#1(#2,#3){\rlap{\calcnum*#1(#2,#3)\relax
\d**=#1\Lengthunit\ifdim\d**<0pt\d**-\d**\fi
\divide\N*2\multiply\N*2\advance\N*1\multiply\N*2\relax
\divide\d**\N*\sm*\n*\*two\sh*(#2,#3){\sl*}\loop
\advance\n*\*one\sh*(#2,#3){\kern-1.48pt\lower.5pt\hbox{\rm.}}\relax
\advance\n*\*one\sh*(#2,#3){\sm*}\advance\n*\*two
\sh*(#2,#3){\sl*}\ifnum\n*<\N*\repeat}}

\def\shl*(#1,#2)#3{\kern#1#3\lower#2#3\hbox{\unhcopy\spl*}}

\def\trianglin#1(#2,#3){\rlap{\toks0={#2}\toks1={#3}\calcnum*#1(#2,#3)\relax
\dd*=.57\Lengthunit\dd*=#1\dd*\divide\dd*\N*
\divide\dd*\*ths \multiply\dd*\magnitude
\d**=#1\Lengthunit\ifdim\d**<0pt\d**-\d**\fi
\multiply\N*2\divide\d**\N*\sm*\n*\*one\loop
\shl**{\dd*}\dd*-\dd*\advance\n*2\relax
\ifnum\n*<\N*\repeat\n*\N*\shl**{0pt}}}

\def\wavelin#1(#2,#3){\rlap{\toks0={#2}\toks1={#3}\calcnum*#1(#2,#3)\relax
\dd*=.23\Lengthunit\dd*=#1\dd*\divide\dd*\N*
\divide\dd*\*ths \multiply\dd*\magnitude
\d**=#1\Lengthunit\ifdim\d**<0pt\d**-\d**\fi
\multiply\N*4\divide\d**\N*\sm*\n*\*one\loop
\shl**{\dd*}\dt*=1.3\dd*\advance\n*\*one
\shl**{\dt*}\advance\n*\*one
\shl**{\dd*}\advance\n*\*two
\dd*-\dd*\ifnum\n*<\N*\repeat\n*\N*\shl**{0pt}}}

\def\w*lin(#1,#2){\rlap{\toks0={#1}\toks1={#2}\d**=\Lengthunit\dd*=-.12\d**
\divide\dd*\*ths \multiply\dd*\magnitude
\N*8\divide\d**\N*\sm*\n*\*one\loop
\shl**{\dd*}\dt*=1.3\dd*\advance\n*\*one
\shl**{\dt*}\advance\n*\*one
\shl**{\dd*}\advance\n*\*one
\shl**{0pt}\dd*-\dd*\advance\n*1\ifnum\n*<\N*\repeat}}

\def\l*arc(#1,#2)[#3][#4]{\rlap{\toks0={#1}\toks1={#2}\d**=\Lengthunit
\dd*=#3.037\d**\dd*=#4\dd*\dt*=#3.049\d**\dt*=#4\dt*\ifdim\d**>10mm\relax
\d**=.25\d**\n*\*one\shl**{-\dd*}\n*\*two\shl**{-\dt*}\n*3\relax
\shl**{-\dd*}\n*4\relax\shl**{0pt}\else
\ifdim\d**>5mm\d**=.5\d**\n*\*one\shl**{-\dt*}\n*\*two
\shl**{0pt}\else\n*\*one\shl**{0pt}\fi\fi}}

\def\d*arc(#1,#2)[#3][#4]{\rlap{\toks0={#1}\toks1={#2}\d**=\Lengthunit
\dd*=#3.037\d**\dd*=#4\dd*\d**=.25\d**\sm*\n*\*one\shl**{-\dd*}\relax
\n*3\relax\sh*(#1,#2){\xL*=\xscale\dd*\yL*=\yscale\dd*
\kern#2\xL*\lower#1\yL*\hbox{\sm*}}\n*4\relax\shl**{0pt}}}

\def\shl**#1{\c*=\the\n*\d**\d*=#1\relax
\a*=\the\toks0\c*\b*=\the\toks1\d*\advance\a*-\b*
\b*=\the\toks1\c*\d*=\the\toks0\d*\advance\b*\d*
\a*=\xscale\a*\b*=\yscale\b*
\rx* \the\cos*\a* \tmp* \the\sin*\b* \advance\rx*-\tmp*
\ry* \the\cos*\b* \tmp* \the\sin*\a* \advance\ry*\tmp*
\raise\ry*\rlap{\kern\rx*\unhcopy\spl*}}

\def\wlin*#1(#2,#3)[#4]{\rlap{\toks0={#2}\toks1={#3}\relax
\c*=#1\l*\c*\c*=.01\Lengthunit\m*\c*\divide\l*\m*
\c*=\the\Nhalfperiods5sp\multiply\c*\l*\N*\c*\divide\N*\*ths
\divide\N*2\multiply\N*2\advance\N*\*one
\dd*=.002\Lengthunit\dd*=#4\dd*\multiply\dd*\l*\divide\dd*\N*
\divide\dd*\*ths \multiply\dd*\magnitude
\d**=#1\multiply\N*4\divide\d**\N*\sm*\n*\*one\loop
\shl**{\dd*}\dt*=1.3\dd*\advance\n*\*one
\shl**{\dt*}\advance\n*\*one
\shl**{\dd*}\advance\n*\*two
\dd*-\dd*\ifnum\n*<\N*\repeat\n*\N*\shl**{0pt}}}

\def\wavebox#1{\setbox0\hbox{#1}\relax
\a*=\wd0\advance\a*14pt\b*=\ht0\advance\b*\dp0\advance\b*14pt\relax
\hbox{\kern9pt\relax
\rmov*(0pt,\ht0){\rmov*(-7pt,7pt){\wlin*\a*(1,0)[+]\wlin*\b*(0,-1)[-]}}\relax
\rmov*(\wd0,-\dp0){\rmov*(7pt,-7pt){\wlin*\a*(-1,0)[+]\wlin*\b*(0,1)[-]}}\relax
\box0\kern9pt}}

\def\rectangle#1(#2,#3){\relax
\lin#1(#2,0)\lin#1(0,#3)\mov#1(0,#3){\lin#1(#2,0)}\mov#1(#2,0){\lin#1(0,#3)}}

\def\dashrectangle#1(#2,#3){\dashlin#1(#2,0)\dashlin#1(0,#3)\relax
\mov#1(0,#3){\dashlin#1(#2,0)}\mov#1(#2,0){\dashlin#1(0,#3)}}

\def\waverectangle#1(#2,#3){\L*=#1\Lengthunit\a*=#2\L*\b*=#3\L*
\ifdim\a*<0pt\a*-\a*\def\x*{-1}\else\def\x*{1}\fi
\ifdim\b*<0pt\b*-\b*\def\y*{-1}\else\def\y*{1}\fi
\wlin*\a*(\x*,0)[-]\wlin*\b*(0,\y*)[+]\relax
\mov#1(0,#3){\wlin*\a*(\x*,0)[+]}\mov#1(#2,0){\wlin*\b*(0,\y*)[-]}}

\def\calcparab*{\ifnum\n*>\m*\k*\N*\advance\k*-\n*\else\k*\n*\fi
\a*=\the\k* sp\a*=10\a*\b*\dm*\advance\b*-\a*\k*\b*
\a*=\the\*ths\b*\divide\a*\l*\multiply\a*\k*
\divide\a*\l*\k*\*ths\r*\a*\advance\k*-\r*\dt*=\the\k*\L*}

\def\arcto#1(#2,#3)[#4]{\rlap{\toks0={#2}\toks1={#3}\calcnum*#1(#2,#3)\relax
\dm*=135sp\dm*=#1\dm*\d**=#1\Lengthunit\ifdim\dm*<0pt\dm*-\dm*\fi
\multiply\dm*\r*\a*=.3\dm*\a*=#4\a*\ifdim\a*<0pt\a*-\a*\fi
\advance\dm*\a*\N*\dm*\divide\N*10000\relax
\divide\N*2\multiply\N*2\advance\N*\*one
\L*=-.25\d**\L*=#4\L*\divide\d**\N*\divide\L*\*ths
\m*\N*\divide\m*2\dm*=\the\m*5sp\l*\dm*\sm*\n*\*one\loop
\calcparab*\shl**{-\dt*}\advance\n*1\ifnum\n*<\N*\repeat}}

\def\arrarcto#1(#2,#3)[#4]{\L*=#1\Lengthunit\L*=.54\L*
\arcto#1(#2,#3)[#4]\rmov*(#2\L*,#3\L*){\d*=.457\L*\d*=#4\d*\d**-\d*
\rmov*(#3\d**,#2\d*){\arrow.02(#2,#3)}}}

\def\dasharcto#1(#2,#3)[#4]{\rlap{\toks0={#2}\toks1={#3}\relax
\calcnum*#1(#2,#3)\dm*=\the\N*5sp\a*=.3\dm*\a*=#4\a*\ifdim\a*<0pt\a*-\a*\fi
\advance\dm*\a*\N*\dm*
\divide\N*20\multiply\N*2\advance\N*1\d**=#1\Lengthunit
\L*=-.25\d**\L*=#4\L*\divide\d**\N*\divide\L*\*ths
\m*\N*\divide\m*2\dm*=\the\m*5sp\l*\dm*
\sm*\n*\*one\loop\calcparab*
\shl**{-\dt*}\advance\n*1\ifnum\n*>\N*\else\calcparab*
\sh*(#2,#3){\xL*=#3\dt* \yL*=#2\dt*
\rx* \the\cos*\xL* \tmp* \the\sin*\yL* \advance\rx*\tmp*
\ry* \the\cos*\yL* \tmp* \the\sin*\xL* \advance\ry*-\tmp*
\kern\rx*\lower\ry*\hbox{\sm*}}\fi
\advance\n*1\ifnum\n*<\N*\repeat}}

\def\*shl*#1{\c*=\the\n*\d**\advance\c*#1\a**\d*\dt*\advance\d*#1\b**
\a*=\the\toks0\c*\b*=\the\toks1\d*\advance\a*-\b*
\b*=\the\toks1\c*\d*=\the\toks0\d*\advance\b*\d*
\rx* \the\cos*\a* \tmp* \the\sin*\b* \advance\rx*-\tmp*
\ry* \the\cos*\b* \tmp* \the\sin*\a* \advance\ry*\tmp*
\raise\ry*\rlap{\kern\rx*\unhcopy\spl*}}

\def\calcnormal*#1{\b**=10000sp\a**\b**\k*\n*\advance\k*-\m*
\multiply\a**\k*\divide\a**\m*\a**=#1\a**\ifdim\a**<0pt\a**-\a**\fi
\ifdim\a**>\b**\d*=.96\a**\advance\d*.4\b**
\else\d*=.96\b**\advance\d*.4\a**\fi
\d*=.01\d*\r*\d*\divide\a**\r*\divide\b**\r*
\ifnum\k*<0\a**-\a**\fi\d*=#1\d*\ifdim\d*<0pt\b**-\b**\fi
\k*\a**\a**=\the\k*\dd*\k*\b**\b**=\the\k*\dd*}

\def\wavearcto#1(#2,#3)[#4]{\rlap{\toks0={#2}\toks1={#3}\relax
\calcnum*#1(#2,#3)\c*=\the\N*5sp\a*=.4\c*\a*=#4\a*\ifdim\a*<0pt\a*-\a*\fi
\advance\c*\a*\N*\c*\divide\N*20\multiply\N*2\advance\N*-1\multiply\N*4\relax
\d**=#1\Lengthunit\dd*=.012\d**
\divide\dd*\*ths \multiply\dd*\magnitude
\ifdim\d**<0pt\d**-\d**\fi\L*=.25\d**
\divide\d**\N*\divide\dd*\N*\L*=#4\L*\divide\L*\*ths
\m*\N*\divide\m*2\dm*=\the\m*0sp\l*\dm*
\sm*\n*\*one\loop\calcnormal*{#4}\calcparab*
\*shl*{1}\advance\n*\*one\calcparab*
\*shl*{1.3}\advance\n*\*one\calcparab*
\*shl*{1}\advance\n*2\dd*-\dd*\ifnum\n*<\N*\repeat\n*\N*\shl**{0pt}}}

\def\triangarcto#1(#2,#3)[#4]{\rlap{\toks0={#2}\toks1={#3}\relax
\calcnum*#1(#2,#3)\c*=\the\N*5sp\a*=.4\c*\a*=#4\a*\ifdim\a*<0pt\a*-\a*\fi
\advance\c*\a*\N*\c*\divide\N*20\multiply\N*2\advance\N*-1\multiply\N*2\relax
\d**=#1\Lengthunit\dd*=.012\d**
\divide\dd*\*ths \multiply\dd*\magnitude
\ifdim\d**<0pt\d**-\d**\fi\L*=.25\d**
\divide\d**\N*\divide\dd*\N*\L*=#4\L*\divide\L*\*ths
\m*\N*\divide\m*2\dm*=\the\m*0sp\l*\dm*
\sm*\n*\*one\loop\calcnormal*{#4}\calcparab*
\*shl*{1}\advance\n*2\dd*-\dd*\ifnum\n*<\N*\repeat\n*\N*\shl**{0pt}}}

\def\hr*#1{\L*=\xscale\Lengthunit\ifnum
\angle**=0\clap{\vrule width#1\L* height.1pt}\else
\L*=#1\L*\L*=.5\L*\rmov*(-\L*,0pt){\sm*}\rmov*(\L*,0pt){\sl*}\fi}

\def\shade#1[#2]{\rlap{\Lengthunit=#1\Lengthunit
\special{em:linewidth .001pt}\relax
\mov(0,#2.05){\hr*{.994}}\mov(0,#2.1){\hr*{.980}}\relax
\mov(0,#2.15){\hr*{.953}}\mov(0,#2.2){\hr*{.916}}\relax
\mov(0,#2.25){\hr*{.867}}\mov(0,#2.3){\hr*{.798}}\relax
\mov(0,#2.35){\hr*{.715}}\mov(0,#2.4){\hr*{.603}}\relax
\mov(0,#2.45){\hr*{.435}}\special{em:linewidth \the\linwid*}}}

\def\dshade#1[#2]{\rlap{\special{em:linewidth .001pt}\relax
\Lengthunit=#1\Lengthunit\if#2-\def\t*{+}\else\def\t*{-}\fi
\mov(0,\t*.025){\relax
\mov(0,#2.05){\hr*{.995}}\mov(0,#2.1){\hr*{.988}}\relax
\mov(0,#2.15){\hr*{.969}}\mov(0,#2.2){\hr*{.937}}\relax
\mov(0,#2.25){\hr*{.893}}\mov(0,#2.3){\hr*{.836}}\relax
\mov(0,#2.35){\hr*{.760}}\mov(0,#2.4){\hr*{.662}}\relax
\mov(0,#2.45){\hr*{.531}}\mov(0,#2.5){\hr*{.320}}\relax
\special{em:linewidth \the\linwid*}}}}

\def\vdot{\rlap{\kern-1.9pt\lower1.8pt\hbox{$\scriptstyle\bullet$}}}
\def\vtimes{\rlap{\kern-3pt\lower1.8pt\hbox{$\scriptstyle\times$}}}
\def\vDot{\rlap{\kern-2.3pt\lower2.7pt\hbox{$\bullet$}}}
\def\vTimes{\rlap{\kern-3.6pt\lower2.4pt\hbox{$\times$}}}

\def\arc(#1)[#2,#3]{{\k*=#2\l*=#3\m*=\l*
\advance\m*-6\ifnum\k*>\l*\relax\else
{\rotate(#2)\mov(#1,0){\sm*}}\loop
\ifnum\k*<\m*\advance\k*5{\rotate(\k*)\mov(#1,0){\sl*}}\repeat
{\rotate(#3)\mov(#1,0){\sl*}}\fi}}

\def\dasharc(#1)[#2,#3]{{\k**=#2\n*=#3\advance\n*-1\advance\n*-\k**
\L*=1000sp\L*#1\L* \multiply\L*\n* \multiply\L*\Nhalfperiods
\divide\L*57\N*\L* \divide\N*2000\ifnum\N*=0\N*1\fi
\r*\n*	\divide\r*\N* \ifnum\r*<2\r*2\fi
\m**\r* \divide\m**2 \l**\r* \advance\l**-\m** \N*\n* \divide\N*\r*
\k**\r* \multiply\k**\N* \dn*\n* \advance\dn*-\k** \divide\dn*2\advance\dn*\*one
\r*\l** \divide\r*2\advance\dn*\r* \advance\N*-2\k**#2\relax
\ifnum\l**<6{\rotate(#2)\mov(#1,0){\sm*}}\advance\k**\dn*
{\rotate(\k**)\mov(#1,0){\sl*}}\advance\k**\m**
{\rotate(\k**)\mov(#1,0){\sm*}}\loop
\advance\k**\l**{\rotate(\k**)\mov(#1,0){\sl*}}\advance\k**\m**
{\rotate(\k**)\mov(#1,0){\sm*}}\advance\N*-1\ifnum\N*>0\repeat
{\rotate(#3)\mov(#1,0){\sl*}}\else\advance\k**\dn*
\arc(#1)[#2,\k**]\loop\advance\k**\m** \r*\k**
\advance\k**\l** {\arc(#1)[\r*,\k**]}\relax
\advance\N*-1\ifnum\N*>0\repeat
\advance\k**\m**\arc(#1)[\k**,#3]\fi}}

\def\triangarc#1(#2)[#3,#4]{{\k**=#3\n*=#4\advance\n*-\k**
\L*=1000sp\L*#2\L* \multiply\L*\n* \multiply\L*\Nhalfperiods
\divide\L*57\N*\L* \divide\N*1000\ifnum\N*=0\N*1\fi
\d**=#2\Lengthunit \d*\d** \divide\d*57\multiply\d*\n*
\r*\n*	\divide\r*\N* \ifnum\r*<2\r*2\fi
\m**\r* \divide\m**2 \l**\r* \advance\l**-\m** \N*\n* \divide\N*\r*
\dt*\d* \divide\dt*\N* \dt*.5\dt* \dt*#1\dt*
\divide\dt*1000\multiply\dt*\magnitude
\k**\r* \multiply\k**\N* \dn*\n* \advance\dn*-\k** \divide\dn*2\relax
\r*\l** \divide\r*2\advance\dn*\r* \advance\N*-1\k**#3\relax
{\rotate(#3)\mov(#2,0){\sm*}}\advance\k**\dn*
{\rotate(\k**)\mov(#2,0){\sl*}}\advance\k**-\m**\advance\l**\m**\loop\dt*-\dt*
\d*\d** \advance\d*\dt*
\advance\k**\l**{\rotate(\k**)\rmov*(\d*,0pt){\sl*}}%
\advance\N*-1\ifnum\N*>0\repeat\advance\k**\m**
{\rotate(\k**)\mov(#2,0){\sl*}}{\rotate(#4)\mov(#2,0){\sl*}}}}

\def\wavearc#1(#2)[#3,#4]{{\k**=#3\n*=#4\advance\n*-\k**
\L*=4000sp\L*#2\L* \multiply\L*\n* \multiply\L*\Nhalfperiods
\divide\L*57\N*\L* \divide\N*1000\ifnum\N*=0\N*1\fi
\d**=#2\Lengthunit \d*\d** \divide\d*57\multiply\d*\n*
\r*\n*	\divide\r*\N* \ifnum\r*=0\r*1\fi
\m**\r* \divide\m**2 \l**\r* \advance\l**-\m** \N*\n* \divide\N*\r*
\dt*\d* \divide\dt*\N* \dt*.7\dt* \dt*#1\dt*
\divide\dt*1000\multiply\dt*\magnitude
\k**\r* \multiply\k**\N* \dn*\n* \advance\dn*-\k** \divide\dn*2\relax
\divide\N*4\advance\N*-1\k**#3\relax
{\rotate(#3)\mov(#2,0){\sm*}}\advance\k**\dn*
{\rotate(\k**)\mov(#2,0){\sl*}}\advance\k**-\m**\advance\l**\m**\loop\dt*-\dt*
\d*\d** \advance\d*\dt* \dd*\d** \advance\dd*1.3\dt*
\advance\k**\r*{\rotate(\k**)\rmov*(\d*,0pt){\sl*}}\relax
\advance\k**\r*{\rotate(\k**)\rmov*(\dd*,0pt){\sl*}}\relax
\advance\k**\r*{\rotate(\k**)\rmov*(\d*,0pt){\sl*}}\relax
\advance\k**\r*
\advance\N*-1\ifnum\N*>0\repeat\advance\k**\m**
{\rotate(\k**)\mov(#2,0){\sl*}}{\rotate(#4)\mov(#2,0){\sl*}}}}

\def\gmov*#1(#2,#3)#4{\rlap{\L*=#1\Lengthunit
\xL*=#2\L* \yL*=#3\L*
\rx* \gcos*\xL* \tmp* \gsin*\yL* \advance\rx*-\tmp*
\ry* \gcos*\yL* \tmp* \gsin*\xL* \advance\ry*\tmp*
\rx*=\xscale\rx* \ry*=\yscale\ry*
\xL* \the\cos*\rx* \tmp* \the\sin*\ry* \advance\xL*-\tmp*
\yL* \the\cos*\ry* \tmp* \the\sin*\rx* \advance\yL*\tmp*
\kern\xL*\raise\yL*\hbox{#4}}}

\def\rgmov*(#1,#2)#3{\rlap{\xL*#1\yL*#2\relax
\rx* \gcos*\xL* \tmp* \gsin*\yL* \advance\rx*-\tmp*
\ry* \gcos*\yL* \tmp* \gsin*\xL* \advance\ry*\tmp*
\rx*=\xscale\rx* \ry*=\yscale\ry*
\xL* \the\cos*\rx* \tmp* \the\sin*\ry* \advance\xL*-\tmp*
\yL* \the\cos*\ry* \tmp* \the\sin*\rx* \advance\yL*\tmp*
\kern\xL*\raise\yL*\hbox{#3}}}

\def\Earc(#1)[#2,#3][#4,#5]{{\k*=#2\l*=#3\m*=\l*
\advance\m*-6\ifnum\k*>\l*\relax\else\def\xscale{#4}\def\yscale{#5}\relax
{\angle**0\rotate(#2)}\gmov*(#1,0){\sm*}\loop
\ifnum\k*<\m*\advance\k*5\relax
{\angle**0\rotate(\k*)}\gmov*(#1,0){\sl*}\repeat
{\angle**0\rotate(#3)}\gmov*(#1,0){\sl*}\relax
\def\xscale{1}\def\yscale{1}\fi}}

\def\dashEarc(#1)[#2,#3][#4,#5]{{\k**=#2\n*=#3\advance\n*-1\advance\n*-\k**
\L*=1000sp\L*#1\L* \multiply\L*\n* \multiply\L*\Nhalfperiods
\divide\L*57\N*\L* \divide\N*2000\ifnum\N*=0\N*1\fi
\r*\n*	\divide\r*\N* \ifnum\r*<2\r*2\fi
\m**\r* \divide\m**2 \l**\r* \advance\l**-\m** \N*\n* \divide\N*\r*
\k**\r*\multiply\k**\N* \dn*\n* \advance\dn*-\k** \divide\dn*2\advance\dn*\*one
\r*\l** \divide\r*2\advance\dn*\r* \advance\N*-2\k**#2\relax
\ifnum\l**<6\def\xscale{#4}\def\yscale{#5}\relax
{\angle**0\rotate(#2)}\gmov*(#1,0){\sm*}\advance\k**\dn*
{\angle**0\rotate(\k**)}\gmov*(#1,0){\sl*}\advance\k**\m**
{\angle**0\rotate(\k**)}\gmov*(#1,0){\sm*}\loop
\advance\k**\l**{\angle**0\rotate(\k**)}\gmov*(#1,0){\sl*}\advance\k**\m**
{\angle**0\rotate(\k**)}\gmov*(#1,0){\sm*}\advance\N*-1\ifnum\N*>0\repeat
{\angle**0\rotate(#3)}\gmov*(#1,0){\sl*}\def\xscale{1}\def\yscale{1}\else
\advance\k**\dn* \Earc(#1)[#2,\k**][#4,#5]\loop\advance\k**\m** \r*\k**
\advance\k**\l** {\Earc(#1)[\r*,\k**][#4,#5]}\relax
\advance\N*-1\ifnum\N*>0\repeat
\advance\k**\m**\Earc(#1)[\k**,#3][#4,#5]\fi}}

\def\triangEarc#1(#2)[#3,#4][#5,#6]{{\k**=#3\n*=#4\advance\n*-\k**
\L*=1000sp\L*#2\L* \multiply\L*\n* \multiply\L*\Nhalfperiods
\divide\L*57\N*\L* \divide\N*1000\ifnum\N*=0\N*1\fi
\d**=#2\Lengthunit \d*\d** \divide\d*57\multiply\d*\n*
\r*\n*	\divide\r*\N* \ifnum\r*<2\r*2\fi
\m**\r* \divide\m**2 \l**\r* \advance\l**-\m** \N*\n* \divide\N*\r*
\dt*\d* \divide\dt*\N* \dt*.5\dt* \dt*#1\dt*
\divide\dt*1000\multiply\dt*\magnitude
\k**\r* \multiply\k**\N* \dn*\n* \advance\dn*-\k** \divide\dn*2\relax
\r*\l** \divide\r*2\advance\dn*\r* \advance\N*-1\k**#3\relax
\def\xscale{#5}\def\yscale{#6}\relax
{\angle**0\rotate(#3)}\gmov*(#2,0){\sm*}\advance\k**\dn*
{\angle**0\rotate(\k**)}\gmov*(#2,0){\sl*}\advance\k**-\m**
\advance\l**\m**\loop\dt*-\dt* \d*\d** \advance\d*\dt*
\advance\k**\l**{\angle**0\rotate(\k**)}\rgmov*(\d*,0pt){\sl*}\relax
\advance\N*-1\ifnum\N*>0\repeat\advance\k**\m**
{\angle**0\rotate(\k**)}\gmov*(#2,0){\sl*}\relax
{\angle**0\rotate(#4)}\gmov*(#2,0){\sl*}\def\xscale{1}\def\yscale{1}}}

\def\waveEarc#1(#2)[#3,#4][#5,#6]{{\k**=#3\n*=#4\advance\n*-\k**
\L*=4000sp\L*#2\L* \multiply\L*\n* \multiply\L*\Nhalfperiods
\divide\L*57\N*\L* \divide\N*1000\ifnum\N*=0\N*1\fi
\d**=#2\Lengthunit \d*\d** \divide\d*57\multiply\d*\n*
\r*\n*	\divide\r*\N* \ifnum\r*=0\r*1\fi
\m**\r* \divide\m**2 \l**\r* \advance\l**-\m** \N*\n* \divide\N*\r*
\dt*\d* \divide\dt*\N* \dt*.7\dt* \dt*#1\dt*
\divide\dt*1000\multiply\dt*\magnitude
\k**\r* \multiply\k**\N* \dn*\n* \advance\dn*-\k** \divide\dn*2\relax
\divide\N*4\advance\N*-1\k**#3\def\xscale{#5}\def\yscale{#6}\relax
{\angle**0\rotate(#3)}\gmov*(#2,0){\sm*}\advance\k**\dn*
{\angle**0\rotate(\k**)}\gmov*(#2,0){\sl*}\advance\k**-\m**
\advance\l**\m**\loop\dt*-\dt*
\d*\d** \advance\d*\dt* \dd*\d** \advance\dd*1.3\dt*
\advance\k**\r*{\angle**0\rotate(\k**)}\rgmov*(\d*,0pt){\sl*}\relax
\advance\k**\r*{\angle**0\rotate(\k**)}\rgmov*(\dd*,0pt){\sl*}\relax
\advance\k**\r*{\angle**0\rotate(\k**)}\rgmov*(\d*,0pt){\sl*}\relax
\advance\k**\r*
\advance\N*-1\ifnum\N*>0\repeat\advance\k**\m**
{\angle**0\rotate(\k**)}\gmov*(#2,0){\sl*}\relax
{\angle**0\rotate(#4)}\gmov*(#2,0){\sl*}\def\xscale{1}\def\yscale{1}}}

\newcount\CatcodeOfAtSign
\CatcodeOfAtSign=\the\catcode`\@
\catcode`\@=11
\def\@arc#1[#2][#3]{\rlap{\Lengthunit=#1\Lengthunit
\sm*\l*arc(#2.1914,#3.0381)[#2][#3]\relax
\mov(#2.1914,#3.0381){\l*arc(#2.1622,#3.1084)[#2][#3]}\relax
\mov(#2.3536,#3.1465){\l*arc(#2.1084,#3.1622)[#2][#3]}\relax
\mov(#2.4619,#3.3086){\l*arc(#2.0381,#3.1914)[#2][#3]}}}

\def\dash@arc#1[#2][#3]{\rlap{\Lengthunit=#1\Lengthunit
\d*arc(#2.1914,#3.0381)[#2][#3]\relax
\mov(#2.1914,#3.0381){\d*arc(#2.1622,#3.1084)[#2][#3]}\relax
\mov(#2.3536,#3.1465){\d*arc(#2.1084,#3.1622)[#2][#3]}\relax
\mov(#2.4619,#3.3086){\d*arc(#2.0381,#3.1914)[#2][#3]}}}

\def\wave@arc#1[#2][#3]{\rlap{\Lengthunit=#1\Lengthunit
\w*lin(#2.1914,#3.0381)\relax
\mov(#2.1914,#3.0381){\w*lin(#2.1622,#3.1084)}\relax
\mov(#2.3536,#3.1465){\w*lin(#2.1084,#3.1622)}\relax
\mov(#2.4619,#3.3086){\w*lin(#2.0381,#3.1914)}}}

\def\bezier#1(#2,#3)(#4,#5)(#6,#7){\N*#1\l*\N* \advance\l*\*one
\d* #4\Lengthunit \advance\d* -#2\Lengthunit \multiply\d* \*two
\b* #6\Lengthunit \advance\b* -#2\Lengthunit
\advance\b*-\d* \divide\b*\N*
\d** #5\Lengthunit \advance\d** -#3\Lengthunit \multiply\d** \*two
\b** #7\Lengthunit \advance\b** -#3\Lengthunit
\advance\b** -\d** \divide\b**\N*
\mov(#2,#3){\sm*{\loop\ifnum\m*<\l*
\a*\m*\b* \advance\a*\d* \divide\a*\N* \multiply\a*\m*
\a**\m*\b** \advance\a**\d** \divide\a**\N* \multiply\a**\m*
\rmov*(\a*,\a**){\unhcopy\spl*}\advance\m*\*one\repeat}}}

\catcode`\*=12

\newcount\n@ast
\def\n@ast@#1{\n@ast0\relax\get@ast@#1\end}
\def\get@ast@#1{\ifx#1\end\let\next\relax\else
\ifx#1*\advance\n@ast1\fi\let\next\get@ast@\fi\next}

\newif\if@up \newif\if@dwn
\def\up@down@#1{\@upfalse\@dwnfalse
\if#1u\@uptrue\fi\if#1U\@uptrue\fi\if#1+\@uptrue\fi
\if#1d\@dwntrue\fi\if#1D\@dwntrue\fi\if#1-\@dwntrue\fi}

\def\halfcirc#1(#2)[#3]{{\Lengthunit=#2\Lengthunit\up@down@{#3}\relax
\if@up\mov(0,.5){\@arc[-][-]\@arc[+][-]}\fi
\if@dwn\mov(0,-.5){\@arc[-][+]\@arc[+][+]}\fi
\def\lft{\mov(0,.5){\@arc[-][-]}\mov(0,-.5){\@arc[-][+]}}\relax
\def\rght{\mov(0,.5){\@arc[+][-]}\mov(0,-.5){\@arc[+][+]}}\relax
\if#3l\lft\fi\if#3L\lft\fi\if#3r\rght\fi\if#3R\rght\fi
\n@ast@{#1}\relax
\ifnum\n@ast>0\if@up\shade[+]\fi\if@dwn\shade[-]\fi\fi
\ifnum\n@ast>1\if@up\dshade[+]\fi\if@dwn\dshade[-]\fi\fi}}

\def\halfdashcirc(#1)[#2]{{\Lengthunit=#1\Lengthunit\up@down@{#2}\relax
\if@up\mov(0,.5){\dash@arc[-][-]\dash@arc[+][-]}\fi
\if@dwn\mov(0,-.5){\dash@arc[-][+]\dash@arc[+][+]}\fi
\def\lft{\mov(0,.5){\dash@arc[-][-]}\mov(0,-.5){\dash@arc[-][+]}}\relax
\def\rght{\mov(0,.5){\dash@arc[+][-]}\mov(0,-.5){\dash@arc[+][+]}}\relax
\if#2l\lft\fi\if#2L\lft\fi\if#2r\rght\fi\if#2R\rght\fi}}

\def\halfwavecirc(#1)[#2]{{\Lengthunit=#1\Lengthunit\up@down@{#2}\relax
\if@up\mov(0,.5){\wave@arc[-][-]\wave@arc[+][-]}\fi
\if@dwn\mov(0,-.5){\wave@arc[-][+]\wave@arc[+][+]}\fi
\def\lft{\mov(0,.5){\wave@arc[-][-]}\mov(0,-.5){\wave@arc[-][+]}}\relax
\def\rght{\mov(0,.5){\wave@arc[+][-]}\mov(0,-.5){\wave@arc[+][+]}}\relax
\if#2l\lft\fi\if#2L\lft\fi\if#2r\rght\fi\if#2R\rght\fi}}

\catcode`\*=11

\def\Circle#1(#2){\halfcirc#1(#2)[u]\halfcirc#1(#2)[d]\n@ast@{#1}\relax
\ifnum\n@ast>0\L*=\xscale\Lengthunit
\ifnum\angle**=0\clap{\vrule width#2\L* height.1pt}\else
\L*=#2\L*\L*=.5\L*\special{em:linewidth .001pt}\relax
\rmov*(-\L*,0pt){\sm*}\rmov*(\L*,0pt){\sl*}\relax
\special{em:linewidth \the\linwid*}\fi\fi}

\catcode`\*=12

\def\wavecirc(#1){\halfwavecirc(#1)[u]\halfwavecirc(#1)[d]}

\def\dashcirc(#1){\halfdashcirc(#1)[u]\halfdashcirc(#1)[d]}

\def\xscale{1}
\def\yscale{1}

\def\Ellipse#1(#2)[#3,#4]{\def\xscale{#3}\def\yscale{#4}\relax
\Circle#1(#2)\def\xscale{1}\def\yscale{1}}

\def\dashEllipse(#1)[#2,#3]{\def\xscale{#2}\def\yscale{#3}\relax
\dashcirc(#1)\def\xscale{1}\def\yscale{1}}

\def\waveEllipse(#1)[#2,#3]{\def\xscale{#2}\def\yscale{#3}\relax
\wavecirc(#1)\def\xscale{1}\def\yscale{1}}

\def\halfEllipse#1(#2)[#3][#4,#5]{\def\xscale{#4}\def\yscale{#5}\relax
\halfcirc#1(#2)[#3]\def\xscale{1}\def\yscale{1}}

\def\halfdashEllipse(#1)[#2][#3,#4]{\def\xscale{#3}\def\yscale{#4}\relax
\halfdashcirc(#1)[#2]\def\xscale{1}\def\yscale{1}}

\def\halfwaveEllipse(#1)[#2][#3,#4]{\def\xscale{#3}\def\yscale{#4}\relax
\halfwavecirc(#1)[#2]\def\xscale{1}\def\yscale{1}}

\catcode`\@=\the\CatcodeOfAtSign

\begin{flushright}
{\bf Preprint SSU-HEP-00/01\\
Samara State University}
\end{flushright}
\vspace{30mm}
\begin{center}
{\bf SELF-ENERGY $O(\alpha^2)$ CORRECTION \\TO THE POSITRONIUM
DECAY RATE}\\

\vspace{4mm}

R.N.~Faustov \\Scientific Council "Cybernetics" RAS\\
117333, Moscow, Vavilov, 40, Russia,\\
A.P.~Martynenko\\ Department of Theoretical Physics, Samara State University,\\
443011, Samara, Pavlov, 1, Russia
\end{center}  

\begin{abstract}
  Self-energy corrections of order O($\alpha^2$) to the
  parapositronium and orthopositronium decay rates are
  calculated. Numerical values of the corresponding
  coefficients are $B_p=-3.74$, $B_o=2.02$.
\end{abstract}

\newpage

Investigation of the parapositronium and orthopositronium decay rates
is one of the important tasks in the QED bound state problem. Especially
interesting experimental situation came into existence about orthopositronium
decay width \cite{KY,KM,KMM,LLM}. While the experiments performed by the University of
Michigan group \cite{We,Niko} lead to strong disagreement between theoretical
and experimental values of orthopositronium decay rate (6-9 standard deviations),
in the experiments of Tokyo University group this discrepancy keep
within the experimental accuracy \cite{Asai}. The theoretical expression for the
orthopositronium decay rate, known at present, can be written in the
form:
\begin{equation}
\Gamma^{th}(o-Ps)=\Gamma_0\left[1-A\frac{\alpha}{\pi}-\frac{\alpha^2}{3}\ln\frac{1}
{\alpha}+B_o\left(\frac{\alpha}{\pi}\right)^2-\frac{3\alpha^3}{2\pi}\ln^2\frac{1}
{\alpha}+...\right],
\end{equation}
\begin{displaymath}
\Gamma_0=\frac{2(\pi^2-9)m\alpha^6}{9\pi},~~~A=-10.286606 (10).
\end{displaymath}
Numerical value of (1) without $\alpha^2$ correction is equal to 7.0382 $\mu s^{-1}$.
Experimental results for the orthopositronium lifetimes are:
\begin{equation}
\Gamma^{exp} (o-Ps,~~~{\rm gas~ measurement})=7.0514(14)~\mu s^{-1}~~\cite{We},
\end{equation}
\begin{displaymath}
\Gamma^{exp} (o-Ps,~~~{\rm vacuum~ measurement})=7.0482(16)~\mu s^{-1}~~\cite{Niko},
\end{displaymath}
\begin{equation}
\Gamma^{exp} (o-Ps,~~~{\rm SiO_2~ measurement})=7.0398(29)~\mu s^{-1}~~\cite{Asai},
\end{equation}
The coefficient $B_o$ should be equal to 41 \cite {CK} so that the expressions (1) and (3)
coincide. Such value of $B_o$ can be derived on the basis of standard
QED calculations \cite{KM,FMS}. During some years the different QED effects
of order O($\alpha^2$) in the $\Gamma^{th}$ were studied \cite{B,BI,AS,AL,Kuraev,AMY},
but exact value of coefficient $B_o$ is not known up to now.
In the case of parapositronium the corrections $\alpha^2$ were calculated
recently in \cite{CMY,CMY1}. The theoretical expression for the decay rate of
singlet positronium is \cite{CMY1}:
\begin{equation}
\Gamma^{th}(p-Ps)=\frac{m\alpha^5}{2}\left[1-\left(5-\frac{\pi^2}{4}\right)
\frac{\alpha}{\pi}+2\alpha^2\ln\frac{1}{\alpha}+B_p\left(\frac{\alpha}{\pi}\right)^2-
\frac{3\alpha^3}{2\pi}\ln^2\frac{1}{\alpha}+...\right],
\end{equation}
where $B_p=1.75(30)$. The corresponding experimental quantity \cite{Al}
\begin{equation}
\Gamma^{exp}(p-Ps)=7990.9\pm 1.7~\mu s^{-1}
\end{equation}
agrees well with the value from (4):
\begin{equation}
\Gamma^{th}(p-Ps)=7989.50 \mu s^{-1}
\end{equation}

Two-loop contributions to the o-Ps lifetime may be devided into several classes:
vertex, vacuum polarization, self-energy and annihilation corrections. Two-loop
corrections due to vacuum polarization were calculated in \cite{BI,AS}. Some
annihilation type contributions to $\Gamma^{th}(o-Ps)$ were studied in
\cite{Kuraev,AMY}. The second order electron self-energy corrections (\cite{YSA,L,P,MPS,Y})
represent the special set of $\alpha^2$ corrections in the decay width of
positronium. Such corrections can be studied independently
by means of renormalized expression for the mass operator.
In this work we calculate $\alpha^2$ corrections
to the positronium decay rate, connected with electron self-energy insertions
to inner electron lines. We don't consider here self-energy vacuum polarization
corrections, which were calculated in \cite{BI,AS}. Corresponding Feynman
diagrams are shown on Fig. 1-2.

The decay width of o-Ps into three photons can be written in the form \cite{FMS,Adkins}:
\begin{equation}
\Gamma(o-Ps\rightarrow 3\gamma)=\frac{1}{3!}\int\frac{d^3k_1}{(2\pi)^3}
\frac{d^3k_2}{(2\pi)^3}\frac{d^3k_3}{(2\pi)^3}(2\pi)^4\delta(P-k_1-k_2-k_3)\times
\end{equation}
\begin{displaymath}
\times\frac{1}{3}\sum_{spin}\sum_{\epsilon_1,\epsilon_2,\epsilon_3}
|T(o-Ps\rightarrow 3\gamma)|^2,
\end{displaymath}
\begin{equation}
T(o-Ps\rightarrow 3\gamma)=\frac{1}{\sqrt{2\omega_1 2\omega_2 2\omega_3 4m}}
M(o-Ps\rightarrow 3\gamma),
\end{equation}
and amplitude $M(o-Ps\rightarrow 3\gamma)$ has the following integral representation:
\begin{equation}
M(o-Ps\rightarrow 3\gamma)=\int\frac{d^3p}{(2\pi)^3}Tr\left[\tilde M(\epsilon_i,\vec p)
\hat\Pi\right]\psi(\vec p),
\end{equation}
where P=(M=2m,0) is the four-momentum of the orthopositronium, $k_1$, $k_2$,
$k_3$ are photon momenta, $\psi(\vec p)$ is the positronium Coulomb wave
function, $\tilde M(\epsilon_i,\vec p)$ is the amplitude of the annihilation
$e^++e^-\rightarrow 3\gamma$, $\hat\Pi$ is the projection operator on the
$^3S_1$ electron-positron state:
\begin{equation}
\hat\Pi=\frac{\hat P+M}{2\sqrt{2}M}\hat\epsilon,
\end{equation}
$\epsilon^\mu$ is the polarization vector of orthopositronium. In the case of
parapositronium we must change $\hat\epsilon\rightarrow\gamma_5$.

\begin{figure}
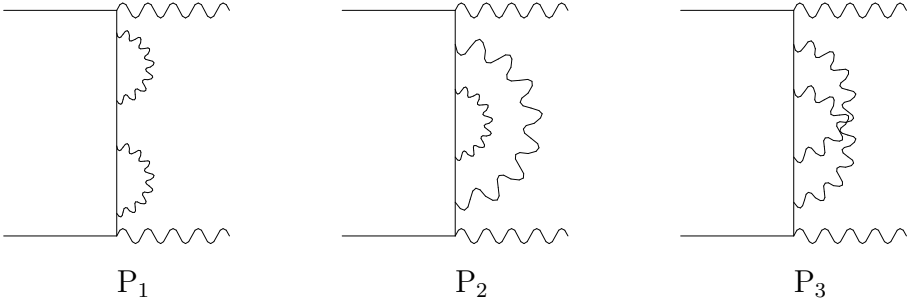

\magnitude=2000
\GRAPH(hsize=15){
\mov(0,0){\lin(1,0)}%
\mov(0,2){\lin(1,0)}%
\mov(1,0){\lin(0,2)}%
\mov(1,0){\wavelin(1,0)}%
\mov(1,2){\wavelin(1,0)}%
\mov(1,0.5){\halfwavecirc(0.6)[R]}%
\mov(1,1.5){\halfwavecirc(0.6)[R]}%
\mov(3,0){\lin(1,0)}%
\mov(3,2){\lin(1,0)}%
\mov(4,0){\lin(0,2)}%
\mov(4,0){\wavelin(1,0)}%
\mov(4,2){\wavelin(1,0)}%
\mov(6,0){\lin(1,0)}%
\mov(6,2){\lin(1,0)}%
\mov(7,0){\lin(0,2)}%
\mov(7,0){\wavelin(1,0)}%
\mov(7,2){\wavelin(1,0)}%
\mov(4,1){\halfwavecirc(1.4)[R]}%
\mov(4,1){\halfwavecirc(0.6)[R]}%
\mov(7,1.2){\halfwavecirc(1.)[R]}%
\mov(7,0.8){\halfwavecirc(1.)[R]}%
\mov(1,-0.5){${\rm P}_1$}%
\mov(4,-0.5){${\rm P}_2$}%
\mov(7,-0.5){${\rm P}_3$}%
}
\caption{Feynman diagrams representing two-loop self-energy
corrections to parapositronium decay rate.}
\end{figure}  

As it is well known the mass operator has no infrared divergences in the
Fried-Yennie gauge (FY) \cite{AP}. So, we can perform the usual mass-shell
subtraction without introduction of the photon mass.
The bare mass operator has the following structure:
\begin{equation}
\Sigma(p)=\delta m+\left(1-Z_2^{-1}\right)(\hat p-m)+\Sigma^{(R)}(p).
\end{equation}
Then the self-energy correction
to the electron line is determined by the substitution:
\begin{equation}
\frac{1}{\hat p-m}\rightarrow\frac{1}{\hat p-m}\Sigma^{(R)}(p)\frac{1}{\hat p-m}.
\end{equation}
The renormalized self-energy operator of the second order in the FY gauge
can be presented as follows \cite{EKS}:
\begin{equation}
\Sigma_{1\gamma}(p)=\frac{\alpha}{4\pi}(\hat p-m)^2\frac{(-3\hat p)}{m^2}\int_0^1dx
\frac{x}{x+\rho(1-x)}=
\end{equation}
\begin{displaymath}
=(\hat p-m)^2\left(-\frac{3\alpha\hat p}{4\pi m^2}\right)\frac{1}{1-\rho}
\left[1+\frac{\rho}{1-\rho}\ln\rho\right],~~~\rho=\frac{m^2-p^2}{m^2}.
\end{displaymath}
We have used this expression to obtain loop-after-loop contributions,
presented by ${\rm P}_1$ and ${\rm O}_1$ graphs.
Feynman diagrams ${\rm P}_2$, ${\rm O}_3$ contain one internal and one
external loops. To find contributions of these diagrams to the mass operator
we take expression (13) for inner loop and again use FY gauge for external photon.
Introducing Feynman parameterization, momentum cutoff and subtracting
ultraviolet divergences in accordance with (11), we have obtain the following
renormalized expression for mass operator with two non-crossed photons:
\begin{equation}
\Sigma^{(R)}_{2\gamma,~a}(p)=\left(\frac{\alpha}{4\pi}\right)^2\Biggl\{\hat p
\Biggl[-9{\rm Li}_2(1-\rho)+\frac{3\pi^2}{2}+\frac{27}{2}-\frac{9}{2(1-\rho)}-
\end{equation}
\begin{displaymath}
-\ln\rho\left(\frac{27}{2}+\frac{9}{2(1-\rho)^2}\right)\Biggr]
+m\left(18\ln\rho-9\right)\Biggr\}.
\end{displaymath}
Explicit expression for the overlapping two-loop self-energy diagram in the
FY gauge was derived in \cite{EKS1}. Corresponding renormalized part of the
mass operator has the form of integral representation over five
Feynman parameters (functions $\sigma_p$, $\sigma_m$):
\begin{equation}
\Sigma^{(R)}_{2\gamma,~b}(p)=\left(\frac{\alpha}{4\pi}\right)^2\frac{(\hat p-m)^2}
{m^2}\left[\hat p\sigma_p(p)+m\sigma_m(p)\right].
\end{equation}
We used this relation to calculate contributions of diagrams ${\rm P}_3$, ${\rm O}_4$.
The results of our calculation of $\alpha^2$ self-energy corrections to the
p-Ps and o-Ps decay rates are shown in Tables 1,2.\\[5mm]
{\bf Table 1. Self-energy $\alpha^2$ corrections to p-Ps decay rate.}\\[3mm]
\begin{tabular}{|c|c|}	\hline
Feynman diagram  &    $B_p$   \\   \hline
${\rm P}_1$   &$\frac{9}{8}\left(1-2\ln 2\right)^2\approx 0.17$ \\   \hline
${\rm P}_2$    & $-\frac{9}{16}\left(\frac{\pi^2}{8}+1-\ln 2\right)\approx -0.87$ \\ \hline
${\rm P}_3$    &-3.04	\\   \hline
\end{tabular}
\vspace{5mm}

\begin{figure}
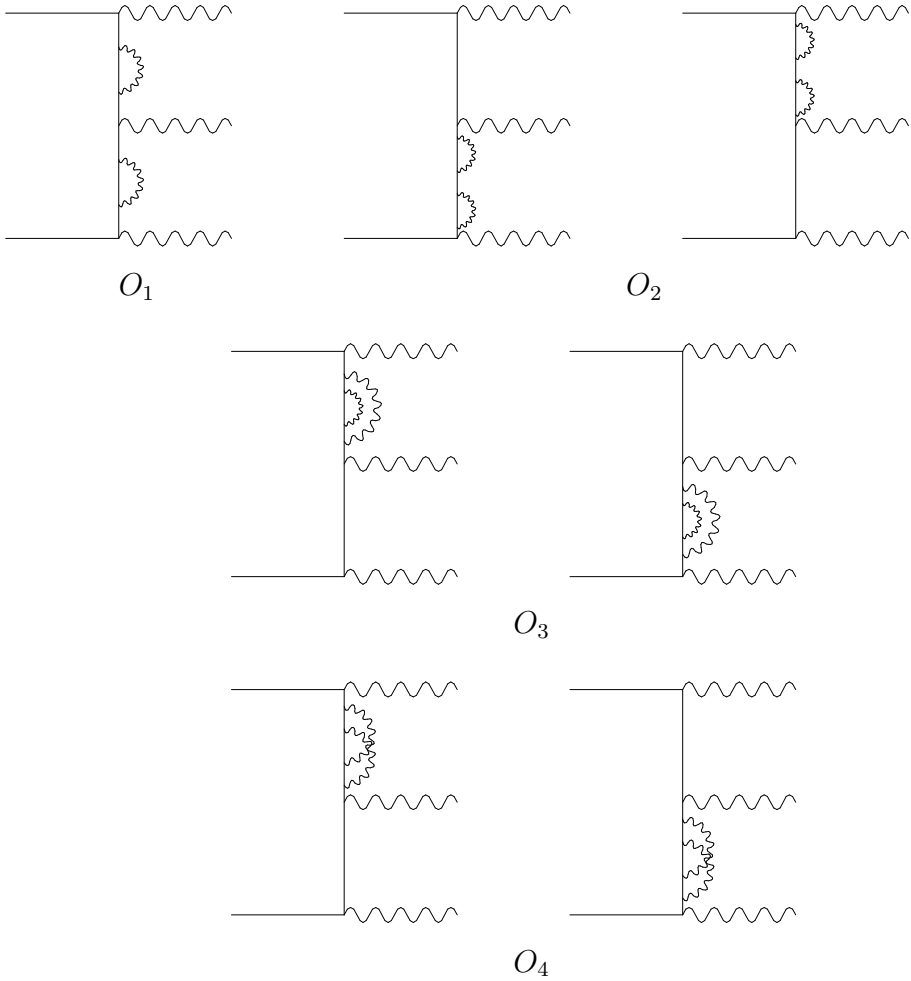

\magnitude=2000
\GRAPH(hsize=15){
\mov(0,0){\lin(1,0)}%
\mov(0,2){\lin(1,0)}%
\mov(1,0){\lin(0,2)}%
\mov(1.,-0.5){$O_1$}%
\mov(5.5,-0.5){$O_2$}%
\mov(4.5,-3.5){$O_3$}%
\mov(4.5,-6.5){$O_4$}%
\mov(1,0.5){\halfwavecirc(0.4)[R]}%
\mov(1,1.5){\halfwavecirc(0.4)[R]}%
\mov(4,0.25){\halfwavecirc(0.3)[R]}%
\mov(4,0.75){\halfwavecirc(0.3)[R]}%
\mov(7,1.25){\halfwavecirc(0.3)[R]}%
\mov(7,1.75){\halfwavecirc(0.3)[R]}%
\mov(1,0){\wavelin(1,0)}%
\mov(1,2){\wavelin(1,0)}%
\mov(1,1){\wavelin(1,0)}%
\mov(3,-1.5){\halfwavecirc(0.6)[R]}%
\mov(3,-1.5){\halfwavecirc(0.3)[R]}%
\mov(6,-2.5){\halfwavecirc(0.6)[R]}%
\mov(6,-2.5){\halfwavecirc(0.3)[R]}%
\mov(3,-4.4){\halfwavecirc(0.5)[R]}%
\mov(3,-4.6){\halfwavecirc(0.5)[R]}%
\mov(6,-5.4){\halfwavecirc(0.5)[R]}%
\mov(6,-5.6){\halfwavecirc(0.5)[R]}%
\mov(3,0){\lin(1,0)}%
\mov(3,2){\lin(1,0)}%
\mov(4,0){\lin(0,2)}%
\mov(4,0){\wavelin(1,0)}%
\mov(4,2){\wavelin(1,0)}%
\mov(4,1){\wavelin(1,0)}%
\mov(6,0){\lin(1,0)}%
\mov(6,2){\lin(1,0)}%
\mov(7,0){\lin(0,2)}%
\mov(7,0){\wavelin(1,0)}%
\mov(7,2){\wavelin(1,0)}%
\mov(7,1){\wavelin(1,0)}%
\mov(2,-1){\lin(1,0)}%
\mov(2,-3){\lin(1,0)}%
\mov(3,-3){\lin(0,2)}%
\mov(3,-1){\wavelin(1,0)}%
\mov(3,-3){\wavelin(1,0)}%
\mov(3,-2){\wavelin(1,0)}%
\mov(2,-4){\lin(1,0)}%
\mov(2,-6){\lin(1,0)}%
\mov(3,-6){\lin(0,2)}%
\mov(3,-4){\wavelin(1,0)}%
\mov(3,-5){\wavelin(1,0)}%
\mov(3,-6){\wavelin(1,0)}%
\mov(5,-4){\lin(1,0)}%
\mov(5,-6){\lin(1,0)}%
\mov(6,-6){\lin(0,2)}%
\mov(6,-4){\wavelin(1,0)}%
\mov(6,-5){\wavelin(1,0)}%
\mov(6,-6){\wavelin(1,0)}%
\mov(5,-1){\lin(1,0)}%
\mov(5,-3){\lin(1,0)}%
\mov(6,-3){\lin(0,2)}%
\mov(6,-1){\wavelin(1,0)}%
\mov(6,-3){\wavelin(1,0)}%
\mov(6,-2){\wavelin(1,0)}%
}
\caption{Feynman diagrams representing two-loop self-energy
corrections to orthopositronium decay rate.}
\end{figure}

Total contributions of the diagrams ${\rm P}_1-{\rm P}_3$, ${\rm O}_1-
{\rm O}_4$ to the coefficients $B_p$, $B_o$ are equal:
\begin{equation}
B^{SE}_p=-3.74,~~~B^{SE}_o=2.02.
\end{equation}
The contribution of the term $\sim B^{SE}_o$ to the o-Ps lifetime increases
the quantity 7.0382 $\mu s^{-1}$ by the value $8\cdot 10^{-5}~\mu s^{-1}$.
Our results for finite part of self-energy contribution to the p-Ps decay
rate do not coincide with the results of \cite{CMY}. We think that the
reason of such difference lies in the choice of gauge for exchanged photons
and in the used renormalization scheme. The calculation of the modulus squared
of the amplitude $|M|^2$ was done by means of the system of analytical
calculations "Form" \cite{FORM}. Many self-energy
corrections for Ps decay width, presented in the Tables were obtained
by the numerical multidimensional integration by means of system "Mathematica"
\cite{W}. Integral representations for the corrections to
the orthopositronium coefficient $B_o$ are given in the Supplement.\\[5mm]
{\bf Table 2. Self-energy $\alpha^2$ corrections to o-Ps decay rate.}\\[3mm]
\begin{tabular}{|c|c|}	\hline
Feynman diagram  &    $B_o$   \\   \hline
${\rm O}_1$   & 0.50   \\   \hline
${\rm O}_2$    &0.20  \\   \hline
${\rm O}_3$    & 5.05 \\   \hline
${\rm O}_4$    & -3.73	\\   \hline
\end{tabular}

\vspace{5mm}

{\large \bf Acknowledgments}

We are grateful to Karshenboim S.G., Khriplovich I.B. for useful
discussions. This work was supported by Russian
Foundation for Basic Research (grant no 98-02-16185), and the Program
"Universities of Russia - Fundamental Researches" (grant no 990192).\\[3mm]

{\large \bf A. Contributions of Feynman diagrams ${\rm O}_1-{\rm O}_4$ to the
coefficient B.}

\begin{equation}
B_{O_1}=\frac{27}{8(\pi^2-9)}\int_0^2d\rho_1\int_{2-\rho_1}^2d\rho_2
\frac{f(\rho_1)f(\rho_2)}{\rho_1\rho_2(4-\rho_1-\rho_2)}\times
\end{equation}
\begin{displaymath}
\times[8-20\rho_1+\frac{1}{2}\rho_1(14\rho_1+15\rho_2)-\frac{1}{4}\rho^2_1
(2\rho_1+5\rho_2)-\frac{1}{4}\rho^2_1\rho_2(\rho_1+2\rho_2)],
\end{displaymath}
\begin{displaymath}
f(\rho_i)=\frac{1}{1-\rho_i}\left(1+\frac{\rho_i}{1-\rho_i}\ln\rho_i\right).
\end{displaymath}
\begin{equation}
B_{O_2}=-\frac{27}{2(\pi^2-9)}\int_0^2d\rho_1\int_{2-\rho_1}^2d\rho_2
\frac{(1-\rho_1+\rho_1\ln\rho_1)^2}{(4-\rho_1-\rho_2)\rho_1\rho_2^2(1-\rho_1)^3}\times
\end{equation}
\begin{displaymath}
\times[16-24\rho_1-10\rho_2+10\rho_1(\rho_1+\rho_2)-\rho_1^3-\frac{7}{4}\rho_1^2
\rho_2+\frac{1}{4}\rho_1\rho_2^2+\frac{1}{4}\rho_2^3-\frac{1}{2}\rho_1\rho_2(
\frac{1}{4}\rho_1^2+\rho_1\rho_2+\frac{1}{4}\rho_2^2)],
\end{displaymath}
\begin{equation}
B_{O_3}=\frac{18}{(\pi^2-9)}\int_0^2d\rho_1\int_{2-\rho_1}^2d\rho_2
\frac{1}{(4-\rho_1-\rho_2)\rho_1^3\rho_2^2}\times
\end{equation}
\begin{displaymath}
\times[s_p(\rho_1)(8-14\rho_1-10\rho_2+\frac{17}{2}\rho_1^2)+\frac{53}{4}\rho_1
\rho_2+4\rho_2^2-\frac{9}{4}\rho_1^3-\frac{39}{8}\rho_1^2\rho_2-\frac{31}{8}\rho_1
\rho_2^2-\frac{1}{2}\rho_2^3+\frac{1}{4}\rho_1^4+
\end{displaymath}
\begin{displaymath}
+\frac{3}{8}\rho_1^3\rho_2+
\frac{7}{16}\rho_1^2\rho_2^2+\frac{3}{8\rho_1}\rho_2^3+\frac{1}{32}\rho_1^4\rho_2+
\frac{1}{8}\rho_1^3\rho_2^2+\frac{1}{32}\rho_1^2\rho_2)+
\end{displaymath}
\begin{displaymath}
+s_m(\rho_1)(8-10\rho_1-10\rho_2+
\frac{7}{2}\rho_1^2+\frac{33}{4}\rho_1\rho_2+4\rho_2^2-\frac{1}{4}\rho_1^3-
\frac{11}{8}\rho_1^2\rho_2-\frac{15}{8}\rho_1\rho_2^2-\frac{1}{2}\rho_2^3-
\end{displaymath}
\begin{displaymath}
-\frac{1}{16}\rho_1^3\rho_2+\frac{1}{8}\rho_1\rho_2^3)],~~
s_p=\frac{3}{4}{\rm Li}_2(1-\rho)-\frac{\pi^2}{8}-\frac{9}{8}+\frac{3}{8(1-\rho)}
+\ln\rho\left(\frac{9}{8}+\frac{3}{8(1-\rho)^2}\right),
\end{displaymath}
\begin{displaymath}
s_m=\frac{3}{4}-\frac{3}{2}\ln\rho.
\end{displaymath}
\begin{equation}
B_{O_4}=-\frac{3}{4(\pi^2-9)}\int_0^2d\rho_1\int_{2-\rho_1}^2d\rho_2
\frac{1}{(4-\rho_1-\rho_2)\rho_1\rho_2^2}\times
\end{equation}
\begin{displaymath}
[\sigma_p(\rho_1)(8-11\rho_1-\frac{15}{2}\rho_2+\frac{9}{2}\rho_1^2+6\rho_1\rho_2+
2\rho_2^2-\frac{1}{2}\rho_1^3-\frac{7}{8}\rho_1^2\rho_2-\frac{3}{8}\rho_1\rho_2^2-
\frac{1}{8}\rho_2^3-\frac{1}{16}\rho_1^3\rho_2-
\end{displaymath}
\begin{displaymath}
-\frac{1}{4}\rho_1^2\rho_2^2-\frac{1}{16}\rho_1\rho_2^3)
+\sigma_m(\rho_1)(\rho_1-\frac{5}{2}\rho_2-\frac{1}{2}\rho_1^2+\rho_1\rho_2+
2\rho_2^2-\frac{1}{2}\rho_1\rho_2^2-\frac{3}{8}\rho_2^3)].
\end{displaymath}


\begin{thebibliography}{99}
\bibitem{KY}Khriplovich I.B., Yelkhovsky A.S. Phys. Lett. {\bf B246}, 520 (1990).
\bibitem{KM}Khriplovich I.B., Milstein A.I. JETF {\bf 79}, 379 (1994).
\bibitem{KMM}Khriplovich I.B., Meshkov I.N., Milstein A.I. JINR Rapid
Communications {\bf N3[83]-97}, (1997).
\bibitem{LLM}Labelle P., Lepage G.P., Magnea U. Phys. Rev. Lett. {\bf 72},
2006 (1994).
\bibitem{We} Westbrook C.I., Gidley D.W., Conti R.S., Rich A. Phys. Rev.
{\bf A40}, 5489 (1989).
\bibitem{Niko}Niko J.S., Gidley D.W., Rich A., Zitzewitz P.W. Phys. Rev. Lett.
{\bf 65}, 1344 (1990).
\bibitem{Asai}Asai S., Orito S., Shinohara N. Phys. Lett. {\bf B357}, 475 (1995).
\bibitem{CK}Czarnecki A., Karshenboim S.G. Decays of Positronium (1999),
hep-ph/9910488.
\bibitem{FMS} Faustov R.N., Martynenko A.P., Saleev V.A. Phys. Rev. {\bf A51},
4520 (1995).
\bibitem{B}Burichenko A.P. Yad. Fiz. {\bf 56}, 123 (1993).
\bibitem{BI}Burichenko A.P., Ivanov D.Yu. Yad. Fiz. {\bf 58}, 898 (1995).
\bibitem{AS}Adkins G.S., Shiferaw Y. Phys. Rev. {\bf A52}, 2442 (1995).
\bibitem{AL}Adkins G.S., Lymberopoulos M. Phys. Rev. {\bf A51}, 2908 (1995).
\bibitem{Kuraev}Antonelli V., Ivanchenko V., Kuraev E., Laliena V. Proc.
Int. Workshop "Hadronic atoms and positronium in the standard model",
Dubna, 26-31 May 1998, p.198.
\bibitem{AMY}Adkins G.S., Melnikov K., Yelkhovsky A. Virtual annihilation
contribution to orthopositronium decay rate (1999), hep-ph/9905553.
\bibitem{CMY}Czarnecki A., Melnikov K., Yelkhovsky A. Phys. Rev. Lett. {\bf 83},
1135 (1999).
\bibitem{CMY1} Czarnecki A., Melnikov K., Yelkhovsky A. $\alpha^2$ corrections
to parapositronium decay: a detailed description (1999), hep-ph/9910488.
\bibitem{Al}Al-Ramadhan A.H., Gidley D.W. Phys. Rev. Lett. {\bf 72}, 1632 (1994).
\bibitem{YSA}Yerokhin V.A., Shabaev V.M., Artemyev A.N. Pis'ma v ZhETF {\bf 66},
19 (1997).
\bibitem{L}Labzowsky L., et al. Phys. Lett. {\bf A240}, 225 (1998).
\bibitem{P}Persson H., et al. Phys. Rev. Lett. {\bf 76}, 1433 (1996).
\bibitem{MPS}Mohr P.J., Plunien G., Soff G. Phys. Rep. {\bf 293}, 227 (1998).
\bibitem{Y}Yerokhin V.A. Loop-after-loop contribution to the second order
Lamb shift in hydrogenlike low-Z atoms (2000), hep-ph/0001327.
\bibitem{Adkins}Adkins G.S. Annals of Physics {\bf 146}, 78 (1983).
\bibitem{AP}Akhiezer A.I., Peletminsky S.V. Fields and Fundamental interactions,
Naukova Dumka, 1986 (in Russian).
\bibitem{EKS}Eides M.I., Karshenboim S.G., Shelyuto V.A. Annals of Physics
{\bf 205}, 231 (1991).
\bibitem{EKS1}Eides M.I., Karshenboim S.G., Shelyuto V.A. Yad. Fiz. {\bf 57},
1309 (1994).
\bibitem{FORM}Vermaseren J.A.M. FORM User's Guide, Amsterdam, 1990.
\bibitem{W}Wolfram S. "Mathematica - A System for Doing Mathematics by
Computer", Addison - Wesley, Reading (MA), 1988.
\end{thebibliography}
\end{document}